\documentclass{article}

\usepackage{arxiv}

\usepackage[utf8]{inputenc} 
\usepackage[T1]{fontenc}    
\usepackage{hyperref}       
\usepackage{url}            
\usepackage{booktabs}       
\usepackage{amsfonts}       
\usepackage{nicefrac}       
\usepackage{microtype}      
\usepackage{lipsum}
\usepackage{graphicx}
\graphicspath{ {./images/} }
\usepackage{soul}
\usepackage{float}
\usepackage{graphicx}
\usepackage{caption}
\usepackage{subcaption}
\usepackage{amsmath}
\usepackage{multirow}

\title{Global Attention based Graph Convolutional Neural Networks for Improved Materials Property Prediction}

\author{
 Steph-Yves Louis$^1$, Yong Zhao$^1$, Alireza Nasiri$^1$, Xiran Wong$^2$, Yuqi Song$^1$, Fei Liu$^{1,3}$, and Jianjun Hu$^{1,3,*}$  \\
 \\
  $^1$Department of Computer Science and Engineering\\
  University of South Carolina\\
  Columbia, SC, 29201\\
   * Correspondence author:
 \texttt{jianjunh@cse.sc.edu}
\And
  \\
  $^2$Department of Statistics\\
  University of South Carolina\\
  Columbia, SC, 29201\\
  \And
  \\
  $^3$School of Mechanical Engineering\\
  Guizhou University\\
  Guiyang, China, 550050\\
}

\begin{document}
\maketitle
\begin{abstract}
Machine learning (ML) methods have gained increasing popularity in exploring and developing new materials. More specifically, graph neural network (GNN) has been applied in predicting material properties. In this work, we develop a novel model, GATGNN, for predicting inorganic material properties based on graph neural networks composed of multiple graph-attention layers (GAT) and a global attention layer. Through the application of the GAT layers, our model can efficiently learn the complex bonds shared among the atoms within each atom's local neighborhood. Subsequently, the global attention layer provides the weight coefficients of each atom in the inorganic crystal material which are used to considerably improve our model's performance. Notably, with the development of our GATGNN model, we show that our method is able to both outperform the previous models' predictions and provide insight into the crystallization of the material.\end{abstract}



\section{INTRODUCTION}

Machine learning and deep learning \cite{krizhevsky2012imagenet,lecun2015deep} have been increasingly used in materials discovery recently with a variety of applications \cite{chen2020critical,agrawal2019deep,vasudevan2019materials} such as rechargeable alkali-Ion batteries, photovoltaics, catalysts, thermoelectrics, superhard materials \cite{mansouri2018machine}, and superconductors \cite{li2020critical}. The two key components of a machine learning model for materials property prediction is the descriptor or feature set and the machine learning algorithm. Currently, there are two main categories of descriptors that are widely used including composition based features \cite{ward2018matminer, cao2019convolutional} and structure based features \cite{kajita2017universal, xie2018crystal,coley2019graph,chen2019graph}. The former has the benefit of being able to be applied to discover new hypothetical materials while the latter has higher prediction performance but only applicable to materials with characterized structure information either experimentally or by computational crystal structure prediction software such as USPEX\cite{oganov2019structure}, which, however, can only predict the structure for relatively simple compositions. 

 
Numerous structural descriptors have been proposed to represent materials such as atom-centered symmetry functions, Coulomb matrix, smooth overlap of atomic positions, deep tensor neural networks, many-body tensor representation, and Voronoi tessellation\cite{chen2020critical}. However, these descriptors all have their own limitations such as not being size-invariant by construction, or basing their representation of infinite crystals on local neighborhoods of atoms in the material \cite{chen2020critical}. Extensive discussions over these structure descriptors and the characteristics of desired structural descriptors such as invariance to translation, rotation, and permutation of homonuclear atoms can be found in recent reviews \cite{chen2020critical,schmidt2019recent}.  



Recently, graph neural networks have gained a lot of attention in materials property prediction \cite{xie2018crystal,chen2019graph,park2019developing} due to high representation learning capabilities and their ability to achieve state-of-the-art results for various problems of classifying graph entities or graph nodes\cite{wu2019comprehensive}. Xie et al. \cite{xie2018crystal} were among the first to apply graph neural networks to materials property prediction by encoding the crystal structures as graphs. Their CGCNN  has achieved strong results in a set of property prediction problems. Notably, the encoding consists of representing the unit cell of the crystal material as a graph such that nodes would represent the  and connecting edges would represent the bonds shared amongst the atoms. A direct benefit of representing the crystal material as a CGCNN-converted graph is the naturally derived vector characterization of the atoms and edges. Chen et al. \cite{chen2019graph} improved the CGCNN model by introducing a global state input including temperature, pressure and entropy. They also found that the element embeddings in their MEGNet models encode periodic chemical trends and can be used for transfer learning based training of models for band gaps and elastic moduli prediction which have limited training data using the embedding learned by models training for formation energy prediction. In another effort to improve CGCNN, Park et al. \cite{park2019developing} proposed an approach to incorporate information of the Voronoi tessellated crystal structure, explicit 3-body correlations of neighboring constituent atoms, and an optimized chemical representation of interatomic bonds in the crystal graphs. 


While previous graph neural network models for crystal materials property prediction has emphasized capturing local atomic environment, here we propose GATGNN, a deep graph neural network model based on global attention mechanism for crystal materials property prediction. The attention mechanism was first introduced into neural networks for natural language processing (NLP) \cite{vasudevan2019materials,zhang2018self}. Basically, attention is simply a vector, often the outputs of dense layer using softmax function, which can be used to learn the contribution of different context vector components to the merged context vector. It has been used to replace the recurrent neural networks and has achieved significant successes in NLP \cite{devlin2018bert, shazeer2020talking,jin2019attention,liu2019deepseqpanii}. In this model, we first use local attention layers to capture properties of local atomic environments and then a global attention layer is used to make weighted aggregation of all these atom environment vectors to create the global representation of the whole crystal structure. This allows our model to better capture the fact that different atoms in the crystal have different contributions to the global material property. In materials science, Coley et. al. \cite{coley2019graph} firstly proposed a global attention mechanism with graph neural network for chemical reactivity prediction. In their model, the global attention  coefficient of each atom is calculated by learning a reaction probability of that atom to all other possible (physically possible) matching . In our work, the coefficient of an atom is calculated by learning either one of 1) its importance based on its location in the graph or 2) its energy contribution to the crystallization of the material. 

The contributions of our work include:
\begin{enumerate}  
\item We propose a global attention mechanism with graph neural network for material's property prediction.
\item Benchmark studies have shown the state-of-the-art performance of our GATNN algorithm in a variety of materials property prediction problems.
\item Ablation experiments have been conducted to demonstrate the advantage of GATGNN.
\item We have extracted physical insights by examining the learned weights of the GATGNN.
\end{enumerate}

\section{METHODS}
\subsection{Data Collections}


The Materials Project database is used to collect the properties of all the crystal materials originally used in the works of CGCNN \cite{xie2018crystal} and MEGNET \cite{chen2020critical}. While the same CGCNN graph-encoding is applied to convert all 46,743 materials from the CGCNN, we modified the graph constructing parameters for about 69,000 MEGNET inorganic materials. We set each node's maximum number of neighbors to 16, the cut-off radius at 4, and the step size for the Gaussian distance at 0.5. 

\subsection{Augmented Graph-Attention Layer}
\begin{figure}[H]
	\centering
	\includegraphics[width=\linewidth]{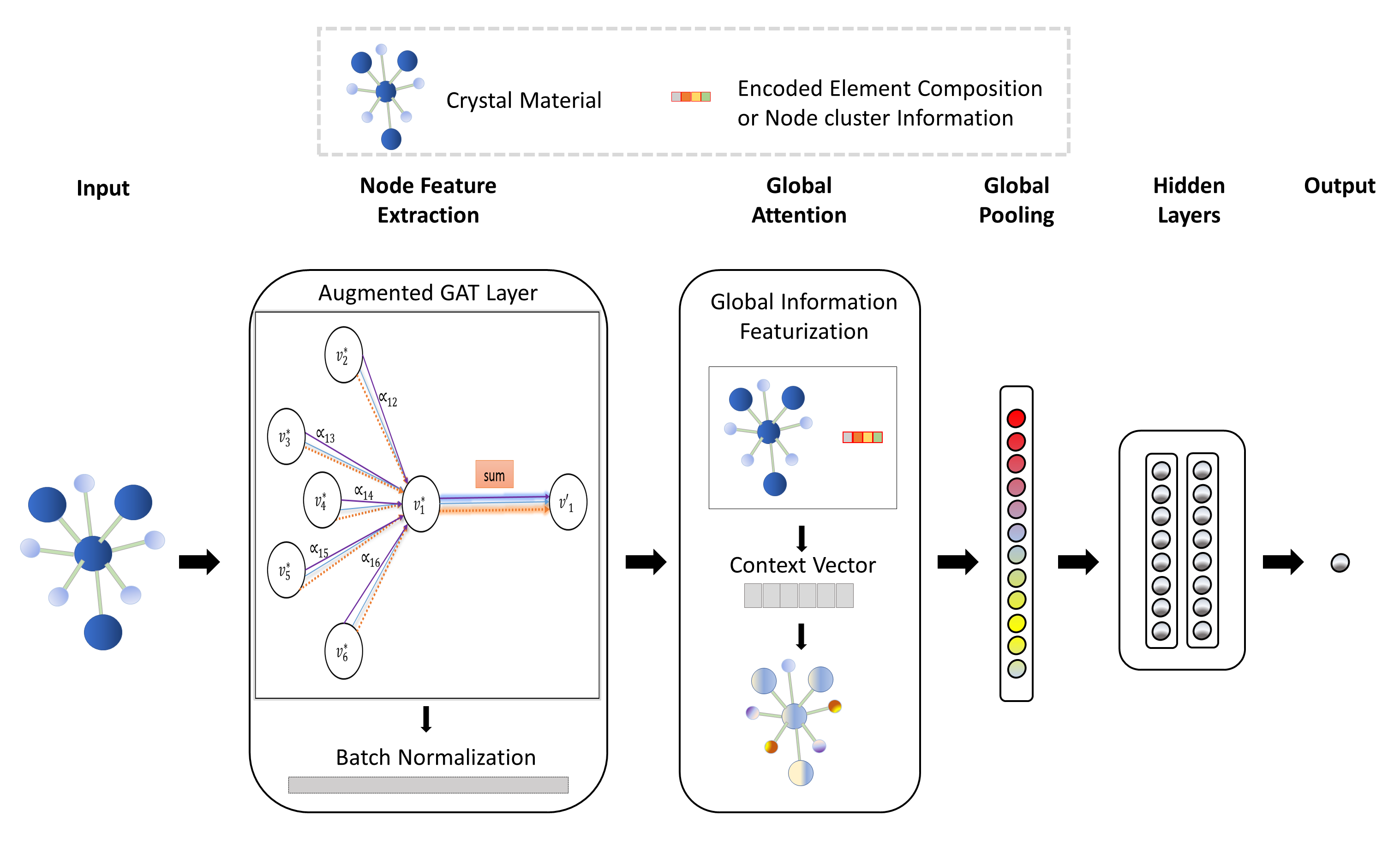}	
  	\caption {Architecture of our global attention graph CNN model GATGNN. Each model is composed of either 5 or 7 of our AGAT layers with 64 neurons. After extracting the node features, a global attention layer is placed before the global pooling of the node feature vectors. Finally, the weighted sum of the crystal feature vector is afterwards fed to two hidden fully-connected layers before a final fully-connected layer outputting the predicted property.}
  	\label{fig:networkarch}
\end{figure}

First, we define a graph as the tuple $(V,E,\textbf{A})$ where $V$ defines a set of nodes, $E$ a set of edges, and $\textbf{A}$ the graph's adjacency matrix. Letting $v \in V$ and $e \in E$ be the feature vectors of atoms and edges, we explain the core of our proposed architecture that is based on the Graph Attention (GAT) Network proposed in 2018 by Velickovice et al. \cite{velivckovic2017graph}. Introduced as a flexible way to adapt the attention mechanism in graph neural networks, GAT Networks proposed a viable solution for dealing with unequal importance of neighboring nodes in graph networks. Making use of the attention heads and context vectors, the GAT architecture makes it possible to learn the importance of the information obtained from neighboring nodes. Notably, the GAT network allows each node to weigh the information from each neighboring node and the learning of these weights can be described by the following operation:
\begin{equation}
\alpha_{ij} ^{(k)}=softmax(g(a_{ij}^{(k)}))=softmax(g(\textbf{a}^T[\textbf{W}^{(k)}\textbf{v}_{i}^{(k-1)}\parallel\textbf{W}^{(k)}\textbf{v}_{j}^{(k-1)}]))
\end{equation}
where $\textbf{a}$ and $\textbf{W}$ define the weight vector and weight matrix of layer $k$, while $\textbf{v}$ and function $g(\cdot)$ define a node  feature vector  and a  non-linear function. For two connected nodes $i$ and $j$, a weight $\alpha_{ij}$ is calculated after applying the softmax function to the attention coefficient $a_{ij}$ of their connection. The attention coefficient $a_{ij}$ is learned via linear transformation of the newly transformed feature vectors of nodes $i$ and $j$ combined by $\textbf{a}$. Once the attention weights are learned, the updated node feature vector is calculated by the following convolution,
\begin{equation}
 \textbf{v}_{i}^{(k)}=\sigma\left(\sum_{j \in  \textit{N}(i)\cup j}  \alpha_{ij} ^{(k)}\textbf{W}^{(k)}\textbf{v}_{j}^{(k-1)}\right)
\end{equation}
, where $\textit{N}(i)$ defines the neighborhood of node $i$ and $\sigma$ the sigmoid function. While GAT networks use the information characterizing neighboring nodes, the original architecture has the limitation of not using any available edge information when learning the attention weights. As a solution to the latter shortcoming, we augmented the original GAT layer. Particularly, we improved the learning capability of the GAT layer by adding Batch-Normalization layers and augmenting the node features vectors with the information from their connecting edge as seen below,
\begin{equation}
\textbf{v*}_{i}^{(k)} = \textbf{v}_{i}^{(k)}\parallel\textbf{e}_{ij}
\end{equation}
where $\textbf{v*}_{i}^{(k)}$ defines the augmented feature vector of node $i$ found by combining $\textbf{v}_i$ and $\textbf{e}_{ij}$, the feature vector of the edge connecting node $i$ and its neighbor $j$. Henceforth, our adaptation of the GAT network for our properties prediction comprises multiple augmented-GAT (AGAT) layers, a global feature pooling layer, and 2 hidden layers. Compared to the properties prediction obtained from applying the original layer, the results of our AGAT layer improved the prediction performance scores by an average of 20\%. By using the batch-normalization, softplus non-linear activation function, and the average of 4 attention heads, we further optimized the learning capability of the AGAT layer. Nonetheless, just the application of the AGAT layers to extract the node features resulted in much poorer predictive performance than previous models from CGCNN and MEGNET works. Certainly, graph neural networks provide strong performance when the main goal is the relationship extraction in a local neighborhood. However, graph neural networks do not use the position of the nodes within the entire graph. Corresponding to the theoretical work, the location of an atom with respect to the structure of the material is critically important and not just to the atoms within the vicinity of its site. Inspired by the works of the [reference below], we propose a global attention layer, before the pooling layer as seen in Figure\ref{fig:networkarch}, that is adapted to particularly translate the information directly learned from a local neighborhood to information at the graph level with meaningful interpretability. 

\subsection{Crystal Global Attention mechanism}
In this work, two types of global information (GI) are used as additional inputs to construct the Global Attention Layer: a feature vector characterizing the entire crystal graph and a feature vector denoting the crystal node's location in the graph. To represent the feature vector characterizing the entire graph, we used the crystal's elemental composition $E$. On the other hand, we represent node i's location in the graph by its cluster $a$: $C_{a}^{i}$, obtained by feeding its coordinates either the Spectral-Clustering or the K-Means algorithm. In our experiments, only three clusters were used, which shows the best performance in our experiments. Then, the global information is propagated to the node feature vectors $\textbf{v}_{i}^{(p)}$ output by the last AGAT layer as defined in table \ref{tab:table1}. Depending on the choice of the global information, the GI method of propagation differs on either concatenation or deconvolution. 

\begin{table}[H]  
\caption{Description of the global information used and the different propagation method throughout the nodes.}
\label{tab:table1}
\begin{center}
\begin{tabular}{|c|p{3.5cm}|l|p{1.5cm}|}
\toprule
\textbf{Global Information (GI)} & \textbf{GI Propagation} & \textbf{Formulation}&\textbf{Method Identifier} \\
\hline
Graph's elemental Composition $E$ & Concatenation with node feature  & $\textbf{v}_{i}^{(p)} = \textbf{v}_{i}^{(p)} \parallel E$&GI M-1\\ 
\hline
\multirow{3}{*}{Node i's Cluster}& Fixed Cluster Unpooling & $\textbf{v}_{i}^{(p)} = \sum_{j \in C_{a}^{i}} \textbf{v}_{j}^{(p)}$ &GI M-2\\ 
\cline{2-4}
& Random Cluster Unpooling & $ \textbf{v}_{i}^{(p)} = {\sum_{j \not{\in}  C_b} \textbf{v}_{j}^{(p)}} \ni C_b \neq  C_{a}^{i}$&GI M-3\\
\cline{2-4}
& Concatenation of graph's pooled feature vector $G$, node features, and one-hot encoding of node clusters &$\textbf{v}_{i}^{(p)} = G \parallel \textbf{v}_{i}^{(p)} \parallel  C_{a}^{i}$&GI M-4\\
\bottomrule


\end{tabular}
\end{center}
\end{table}

For the choice of the crystal's elemental composition $E$, the GI propagation method consists of concatenating $E$ to the feature vector of each node $\textbf{v}_{i}^{(p)}$. Figure 2 provides an illustration of the schema of the propagation method of the elemental composition. For using the node's location as global information, we propose two new strategies. The first strategy defines the concatenation of the pooled feature vector of the crystal graph $G$ (eq. \eqref{eq:4}), the feature vector of each node $\textbf{v}_{i}^{(p)}$, and the one-hot encoded cluster information of each node $C_{a}^{i}$. The combination of the nodes' feature vector and cluster information with G, is then fed to a single feed-forward layer which outputs a new feature vector encoding the node's location. 

\begin{figure}[H]
\centering
\includegraphics[width=\linewidth]{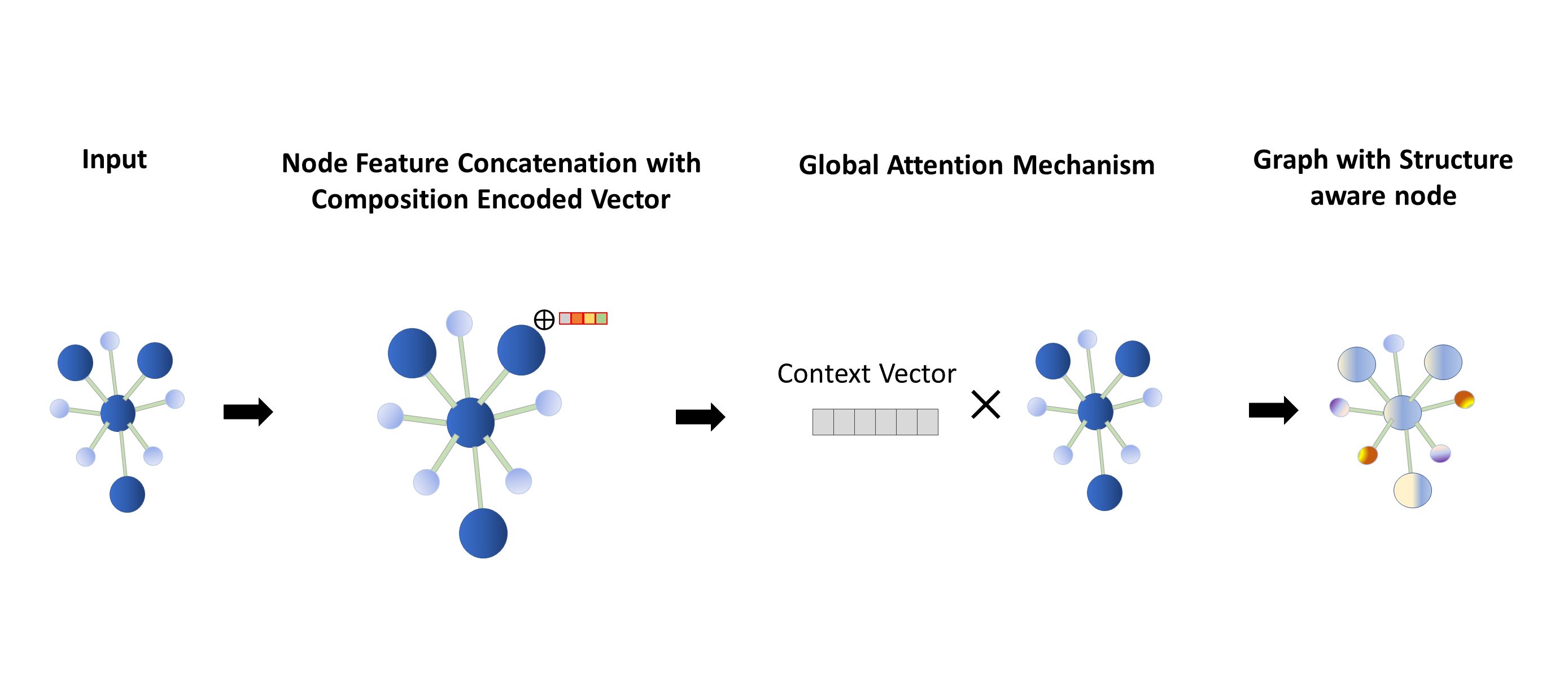}
\caption{Illustration of the global information propagation for the material's elemental composition through the nodes of the crystal. Each crystal node's feature vector is concatenated with the composition encoded vector, which outputs a context vector after forwarded to two fully connected layers }
\end{figure}    
\label{fig:composition}

\begin{equation}
G =\sum_{i=1}^{N}\textbf{v}_{i}^{(p)}
\end{equation}
\label{eq:4}

Our last strategy for propagating each node's position throughout the entire crystal, illustrated in Figure 3, is based on our operations of pooling and unpooling the cluster features of the graph nodes. Upon separating the crystal nodes into clusters, we apply our cluster-pooling operation, which is the summation of the feature vector of all nodes belonging to that specific cluster. The next operation, the cluster unpooling or deconvolution, consists of replacing the feature vectors of each node by cluster-pooled feature vectors. Each node's feature vector can be replaced either randomly or non-randomly (fixed). In the fixed unpooling method, each node's feature vector is replaced by the summed feature vector of the cluster that node belongs to. On the other hand, the random unpooling method defines the method in which each node's feature vector is replaced by the summed feature vector of any cluster that the node did not belong to. Essentially, our proposed global attention layer relies on the distribution of a meaningful and universal information though all the nodes in the crystal graph. Once all of the graph nodes' feature vectors are updated, they are fed to a feed-forward neural network with two fully connected layers and a softmax layer outputting a coefficient vector $\textbf{c}$. Notably, that coefficient vector $\textbf{c}$ contains all the corresponding coefficient $c_i$ for each node $i$ in the graph. Finally, each $c_i$ is then multiplied by the corresponding feature vector $\textbf{v}_{i}^{(p)}$ of the matching node. 

\begin{figure}[H]
\centering
\includegraphics[width=\linewidth]{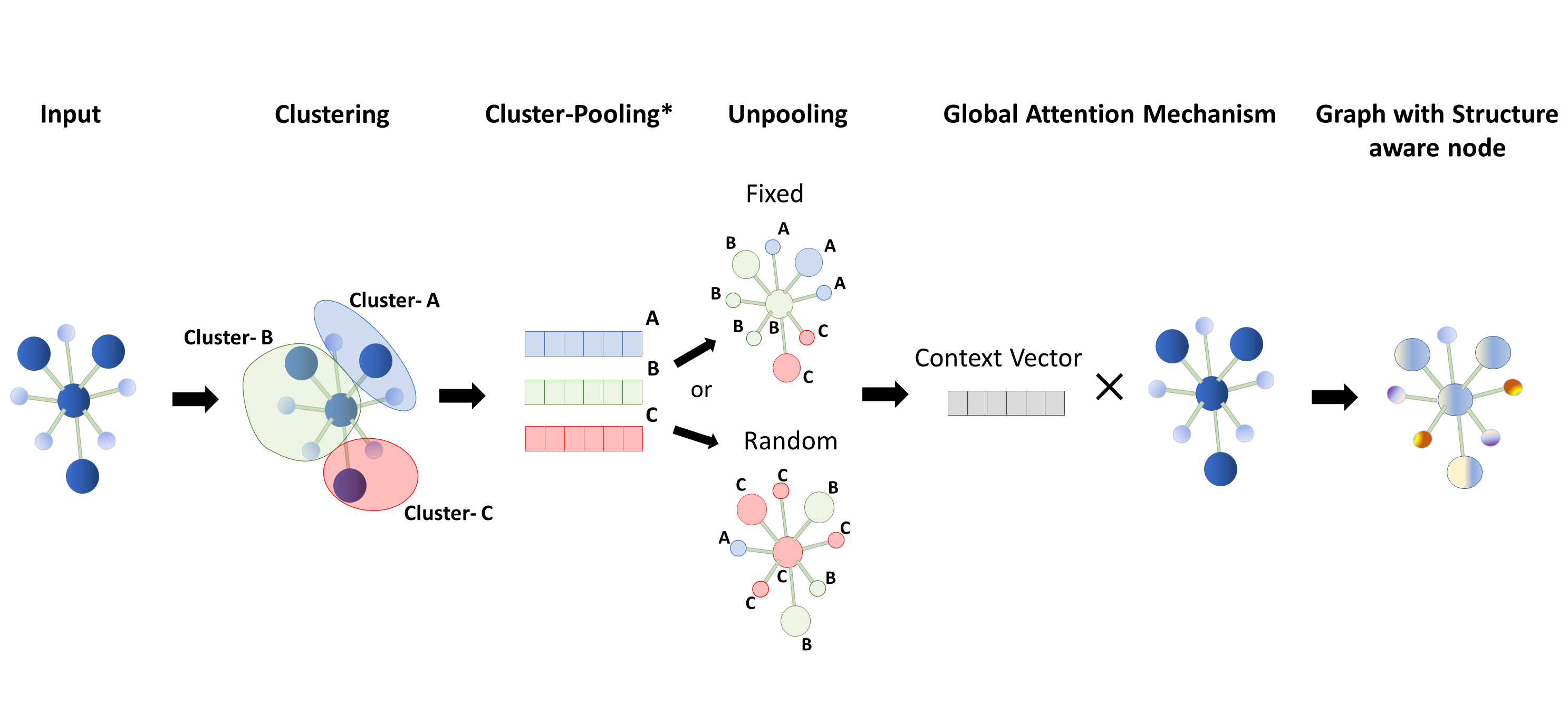}
\caption{Illustration of two propagation methods for the node's cluster. The Cluster-Pooling * defines the aggregation of the feature vectors of all nodes that belong to each of the three clusters. The clusters feature vectors are redistributed through the all the nodes in the unpooling method, which is either fixed or random. After feeding the crystal graph with updated node feature vectors, a context vector is output. The context vector, which contains the weight relating to each node's location, is then multiplied with the input crystal node feature vectors to provide a graph with Structure aware nodes.}
\end{figure}   

\subsection{Model Construction and Training}
To construct our models, we used the open-source library of Pytorch \cite{paszke2017automatic} and its geometric deep learning extension of Pytorch-geometric \cite{fey2019fast}. A model for each designated property was trained for a total of 500 epochs using early-stopping with a patience parameter 150. The additional model hyper-parameters describe a learning rate initiated at $5\cdot 10^{-3}$ which later reduces to $5\cdot 10^{-4}$ and then to $5\cdot 10^{-5}$, a batch-size of 256, 5 or 7 of our AGAT layers with 64 neurons, 4 attention-heads, the Glorot uniform weight initialization, the SmoothL1loss as loss function, and the AdamW optimizer. 

The separation of the data into training, testing, and validation sets were carried as described in the original works \cite{xie2018crystal,chen2019graph}. A split of 60:20:20 of the whole datasets in CGCNN study was used to train and evaluate our models while a split of 80:10:10 was used for training and evaluating our models as is done in MEGNET study. All of the models were trained on Nvidia GeForce RTX 2080 Ti GPUs. During the training process, the model parameters leading to the lowest validation error are saved to construct the final model.

\section{Results}
\subsection{Prediction Performance Improvements over MEGNET and CGCNN}
Table 2 provides a summary of the comparison between our models' performance and the models of CGCNN. From the reported mean absolute error (MAE) scores, our models outperform the CGCNN models in nearly all properties prediction problems. Except for the formation-energy prediction problem for which the scores matched, the absolute amount in performance improvement over the previous CGCNN results range from 2.3\% to 33\%. Material properties with 10\% score improvement were observed including Absolute-Energy, Band-Gap, and Bulk-Moduli as shown in the table.


\begin{table}[H] 
\caption{Performance (MAE) comparison over seven materials property prediction problems of our model compared to CGCNN. The number of training samples used for each model is indicated in parentheses.}
\begin{center}
 \begin{tabular}{||c c c c||} 
  \hline
Properties & CGCNN & GATGNN(this work) & Units\\ [0.5ex] 
 \hline\hline
Formation Energy & 0.039 (28,046) & 0.039 (28,046) & eV/atom\\ 
  \hline
Absolute Energy & 0.072 (28,046) & \textbf{0.048} (28,046) & eV/atom \\
 \hline
Fermi Energy & 0.363 (28,046) & \textbf{0.33} (21,885) & eV/atom\\
\hline
 Band Gap & 0.388 (16,485) & \textbf{0.322} (16,485) & eV\\
 \hline
 Bulk-Moduli & 0.054 (2,041) & \textbf{0.047} (2,041)& log(GPa) \\
 \hline 
 Shear-Moduli & 0.087 (2,041) & \textbf{0.085} (2,041)& log(GPa)  \\ 
 \hline 
 Poisson-ratio & 0.030 (2,041) & \textbf{0.029} (2,041) & -- \\ 
 \hline 
\end{tabular}
\end{center}
\end{table}

Consistent with the performance comparison with CGCNN, our GATGNN models likewise outperform the MEGNET models in the predictions of three out of four properties as shown in Table \ref{table:compareMEGNET}, which displays the side-by-side comparison of the MEGNET results compared to our models. While our models improves the prediction results reported by MEGNET in predicting the Band-Gap, Bulk-Moduli, and Shear-Moduli properties, our models underperformed in the prediction of the Formation Energy property. The lower performance of our model for the Formation Energy prediction compared to those of both CGCNN and MEGNET suggests one potential limitation of our models.

\begin{table}[H] 
\begin{center}
\caption{MAE comparison of four materials properties prediction of our GATGNN model compared to MEGNET. The number of training samples used for each model is indicated in parentheses.}
\label{table:compareMEGNET}
 \begin{tabular}{||c c c c||} 
  \hline
Properties & MEGNET & GATGNN(this Work) & Units \\ [0.5ex] 
 \hline\hline
Formation Energy & \textbf{0.028} $\pm${0.000} (60,000) & 0.048 (60,000) & eV/atom\\  \hline
 Band Gap & 0.33 $\pm$0.01 (36,720) & \textbf{0.31} (36,720) & eV \\
 \hline
 Bulk-Moduli & 0.050 $\pm$0.002 (4664) & \textbf{0.045} (4664) & log(GPa) \\
 \hline 
 Shear-Moduli & 0.079 $\pm$ 0.003 (4664) & \textbf{0.075} (4664)& log(GPa) \\ 
 \hline 
\end{tabular}
\end{center}
\end{table}

\subsection{Ablation experiments}
In Table 4, we provide the MAE comparison for the prediction of the Band-Gap, Bulk-Moduli, and Shear-Moduli properties for all the attention models investigated in this work. Without our minor improvements and the usage of edge information, the original GAT demonstrated a very poor predictive ability for the three tested properties. While the AGAT model recorded comparable results to CGCNN, all of our global information propagation solutions improved its performance with the added attention coefficient vector. For each one of the three properties predictions, the four solutions provided very similar results. The reported descriptive statistics for the Band-Gap, Bulk-Moduli, and Shear-Moduli properties listed as $0.33\pm{0.005}$, $0.0495\pm{0.002}$, and $0.0868\pm{0.0015}$ indicate that the application of the weight coefficients from any GI-propagation method would yield better prediction than without its usage. Even though the third GI-Method of Random Unpooling didn't yield the lowest MAE for either one of the properties, it should be noted that this method showed the most consistency of improving the AGAT performance. In two out of the three properties, the usage of the crystal elemental composition provided the lowest prediction errors. In two out the three properties prediction problems,the learning of the node position through our GI M-4 method of properties yielded the worst result. Figure 4 displays the performance of our model for the GI M-4 method. For none of the properties did the learning of the position with our GI M-4 method of propagation outperform either the fixed or the random unpooling propagation.   
\begin{table}[H]
\begin{center}
\caption{Performance (MAE) comparison over three properties prediction for graph networks with original GAT layer, our AGAT layer, and our four proposed methods of GI propagation.}
\label{table:globalattention}
 \begin{tabular}{|| c c c c c c c c c ||} 
  \hline
Properties &GAT &AGAT & GI M-1& GI M-2& GI M-3& GI M-4 & Units &\# Training Samples \\
 \hline\hline
 Band Gap &0.466&0.345 &0.329&\textbf{0.322}&0.332&0.337& eV & 16,485 \\
 \hline
 Bulk-Moduli & 0.081 & 0.054 &\textbf{0.047}&0.051&0.048 &0.052& log(GPa) & 2,041\\
 \hline 
 Shear-Moduli & 0.121 & 0.094 &\textbf{0.085}&0.089&0.086&0.087& log(GPa) &2,041 \\
 \hline 
\end{tabular}
\end{center}
\end{table}

\section{Discussion}
The architecture of our graph neural network offers the inherent benefits of improved accuracy and interepretability. With the use of either one of our proposed global attention layers for properties prediction, we observed a minimum reduction of prediction error of 7\% for all property prediction problems evaluated in this study including band gap, bulk modulus and Shear modulus as shown in Table\ref{table:globalattention}. 
 Even though our models were trained using a larger number of parameters than CGCNN and a comparable number of parameters to MEGNET, our models have the advantage of a much faster convergence than the latter. For most of the models, about 160 epochs are enough for our GATGNN models to reach comparable results to CGCNN and MEGNET which usually take 500 to 2000 epochs or more with the same batch sizes. Using the bulk-moduli property as an example, Figure 4 displays the resulting training and parity plots from our models. Figure 4 a) provides the details of the predictive performance from our GI M-1 attention model. Our model recorded its lowest validation loss at epoch 252 at which point its parameters were saved. Then, the latter model, with its saved parameters from epoch 252, is later evaluated on the testing set to display the parity plot shown in Figure 4 b) for a MAE of 0.048.  While our reported results come from the performance of the model with five AGAT layers for our CGCNN comparison models and seven for our MEGNET models, using a minimum of three layers with four attention heads would also provide a similar behavior of faster convergence and similar state-of-the-art results.

\begin{figure}[H]
	\centering
	\begin{subfigure}[b]{.45\textwidth}
		\includegraphics[width=\textwidth]{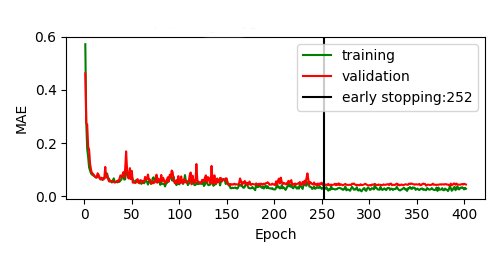}
		\caption{Training plot}
	\end{subfigure}
	\begin{subfigure}[b]{.45\textwidth}
		\includegraphics[width=\textwidth]{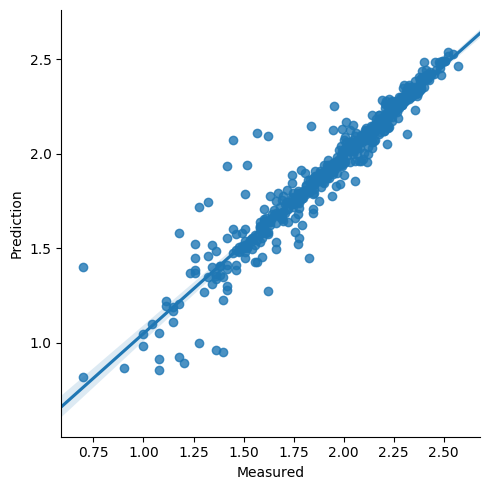}
		\caption{ Parity plot}
	\end{subfigure}
	\caption{Training plot (a) and parity plot (b) for the CGCNN Bulk-Moduli trained model; with property on the log scale.}
	\label{fig:parity}
\end{figure}

To obtain additional insight from our GATGNN model, Table 5 provides a detailed analysis of the extracted context vector from the prediction of the material mp-20452 which is displayed in Figure 5. The table lists the interpretable coefficient corresponding to each atom in the unit cell by comparing the results of our proposed model without global attention to the results of the four global attention mechanisms discussed in this work. The last row in the table corresponds to the predicted bulk-moduli property for the material with formula (Ba(MgPb)$_2$). Except for the model trained without the global attention layer, each global attention method provides a meaningful weight to each node with respect to the global information used. 

\begin{table}[H]
 \begin{center}
 \caption{Context vectors obtained in predicting bulk modulus ($K_{VRH}$) of material mp-20452 (with $K_{VRH} = 1.5051$) obtained from each method discussed in our work. Each row lists the rounded weight contribution of an atom in the unit cell for each proposed method. The last row: Predicted $K_{VRH}$ lists the obtained predicted property from each method. }

 \begin{tabular}{||c c c c c c ||} 
\hline
Atom &AGAT & GI M-1& GI M-2& GI M-3& GI M-4 \\
 \hline\hline
Ba & 1.0&0.353 &0.05 &0.1 &0.185 \\
\hline 
Ba & 1.0&0.353 &0.185 &0.1 &0.185 \\
\hline 
Mg & 1.0&0.056 &0.05 &0.098 &0.12 \\
\hline 
Mg & 1.0&0.056 &0.185 &0.1 &0.12 \\
\hline 
Mg & 1.0&0.017 &0.037 &0.1 &0.033 \\
\hline 
Mg & 1.0&0.017 &0.037 &0.1 &0.033 \\
\hline 
Pb & 1.0&0.06 &0.05 &0.098 &0.124 \\
\hline  
Pb & 1.0&0.06 &0.037 &0.1 &0.124 \\
\hline 
Pb & 1.0&0.013 &0.185 &0.1 &0.037 \\
\hline 
 Pb & 1.0 &0.013 &0.185 &0.1 &0.037 \\
\hline \hline
Predicted $K_{VRH}$&1.5977&\textbf{1.5024}&1.5528&1.5085&1.5204 \\
\hline
\end{tabular} 
 \end{center}
\end{table}

\begin{figure}[H]
\centering
\includegraphics[width=6cm]{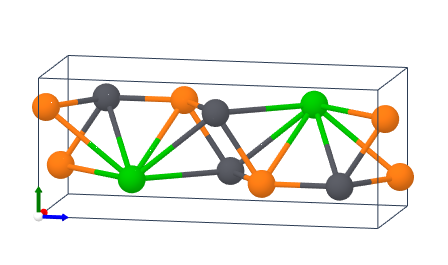}
\caption {Ba(MgPb)$_2$ Crystal structure.  (Material-Project id: mp-20452). Orange nodes represent the Mg sites, green the Ba sites, and grey the Pb sites.}
\label{fig:3D_arch}
\end{figure}

 With all of its weights fixed at 1.0, the AGAT model trained without our global attention had the largest error of 0.09. On the other hand, each one of the other global attention methods returned a mixture of weights per node which offer a different level of information regarding the GI and GI propagation method. The GI M-1 weighted the  by their types. Barium (Ba) accounted for about 70\% of the sum of the context vector while Lead (Pb) and Magnesium (Mg) had a similar weight totaling to 0.146. Particularly, we note that for the four Mg and four Pb atoms, two of each type have half of the weight assigned to the remaining two. The three other global attention methods weight assigned the weights to the nodes based on their location in the graph. 
 
 The GI M-4 had a similar distribution of weights as the GI M-1 method. The global attention layer values the sites of Ba sites as the most critical locations attributing a total weight of 0.37 to its atoms. The Lead and Magnesium  had weights totaling to 0.322 and 0.304. As in GI M-1, half of the Lead  and half of the Magnesium  had much more significant weights than the remaining. The GI M-2 and GI M-3 methods weighted the nodes based on the node clusters of the graph. For the GI M-2 method, weights are attributed to the nodes based on their corresponding clusters. For the three clusters formed for the mp-20542 material, their corresponding weights correspond to 0.185, 0.05, and 0.037. The cluster that contains the second Ba, the first Mg, and the last two Lead atoms has a cumulative weight of 0.74. The next most important cluster, containing the first Ba, Mg, and Pb, has a cumulative weight of 0.15 while the last cluster totals a weight of 0.11. Lastly, the GI M-3 method provides weights on the basis that the nodes belonged to a different site. With the exception of the first Mg and Pb , all of the  are attributed the same weight of 0.1. 
 
 Ultimately, with the application of our graph attention layers from which we can extract the nodes' weights, we are able to derive a more effective representation of the crystal feature vector. After that, our global attention layer transforms each node's feature vector to the graph scale by multiplying it with the learned weight to extract the pooled crystal feature vector. In a set of experiments that we conducted with the Random Forest (RF) algorithm, we noted that the extracted pooled crystal feature could allow RF to achieve lower prediction error than the well-known Magpie features \cite{ward2016general}.
 Essentially, our graph network model allows us to extract the pooled crystal feature as a good representative set of features for that crystal, which can be used for property prediction, classification, or other downstream tasks.
 Furthermore, like MEGNET \cite{park2019developing}, our  GATGNN model also provides the ability to transfer the learned node relationships and weights from a previously built model trained with large amount of labelled data to training models with limited property data.

\section{Conclusions}

We proposed GATGNN, a novel graph convolutional neural network model with global attention mechanisms for accurate materials property prediction. Evaluations on standard materials projects dataset over multiple materials properties such as band gap, bulk-moduli, shear-moduli, formation energy and etc show that our global attention based graph convolutional network models can achieve better performance compared to both CGCNN and its improved MEGNET model. Further analysis shows the importance of global information in accurate prediction of crystal materials properties, which has been sought in both MEGNET and the improved CGCNN \cite{park2019developing}.


\vspace{6pt} 



\section{Author contribution}

Conceptualization, J.H. and S.L.; methodology, S.L. and J.H.; software, S.L.; validation, S.L., Y.Z. and A.N.;  investigation, S.L. and J.H.; resources, J.H.; data curation, S.L. and Y.S. and F.L.; writing--original draft preparation, S.L. and J.H.; writing--review and editing, S.L. and J.H; visualization, S.L. and X.W.; supervision, J.H.;  funding acquisition, J.H.

\section{Funding}

 Research reported in this work was supported in part by NSF under grant and 1940099 and 1905775 and by NSF SC EPSCoR Program under award number (NSF Award OIA-1655740 and GEAR-CRP 19-GC02). The views, perspective, and content do not necessarily represent the official views of the SC EPSCoR Program nor those of the NSF. This work was also partially supported.


\section{Data and code availability}

The datasets are downloaded from Materials Project database. The source code will be are available from the corresponding authors upon reasonable request after official publication of the manuscript.

\bibliographystyle{unsrt}

\end{document}